\documentclass[manuscript]{acmart}
\usepackage{graphicx}
\usepackage{multirow}
\usepackage{booktabs}
\usepackage{tabularx}
\usepackage{hyperref}
\usepackage{url}
\usepackage{multirow}
\usepackage{caption}
\usepackage{subcaption}
\usepackage{xcolor}
\usepackage{color, colortbl}
\usepackage{arydshln}

\definecolor{Gray}{gray}{0.9}

\newcommand{\nemig}{NeMig}
\newcommand{\nemigkg}{NeMigKG}

\AtBeginDocument{%
  \providecommand\BibTeX{{%
    \normalfont B\kern-0.5em{\scshape i\kern-0.25em b}\kern-0.8em\TeX}}}

\setcopyright{acmcopyright}
\copyrightyear{2018}
\acmYear{2018}
\acmDOI{XXXXXXX.XXXXXXX}

\acmConference[INRA 2023]{11th International Workshop on News Recommendation and Analytics}{September 18--22, 2023}{Singapore, Singapore}
\acmPrice{15.00}
\acmISBN{978-1-4503-XXXX-X/18/06}

\begin{document}

\title{NeMig - A Bilingual News Collection and Knowledge Graph about Migration}


\author{Andreea Iana}
\email{andreea.iana@uni-mannheim.de}
\affiliation{%
  \institution{University of Mannheim}
  \country{Germany}
}

\author{Mehwish Alam}
\email{mehwish.alam@telecom-paris.fr}
\affiliation{%
  \institution{Télécom Paris, Institut Polytechique de Paris}
  \country{France}
}

\author{Alexander Grote}
\email{alexander.grote@kit.edu}
\author{Nevena Nikolajevic}
\email{nevena.nikolajevic@kit.edu}
\affiliation{%
  \institution{Karlsruhe Institute of Technology}
  \country{Germany}
}

\author{Katharina Ludwig}
\email{andreea.iana@uni-mannheim.de}
\author{Philipp Müller}
\email{p.mueller@uni-mannheim.de}
\affiliation{%
  \institution{University of Mannheim}
  \country{Germany}
}

\author{Christof Weinhardt}
\email{weinhardt@kit.edu}
\affiliation{%
  \institution{Karlsruhe Institute of Technology}
  \country{Germany}
}

\author{Heiko Paulheim}
\email{heiko.paulheim@uni-mannheim.de}
\affiliation{%
  \institution{University of Mannheim}
  \country{Germany}
}

\renewcommand{\shortauthors}{Iana, et al.}

\begin{abstract}
  News recommendation plays a critical role in shaping the public's worldviews through the way in which it filters and disseminates information about different topics. Given the crucial impact that media plays in opinion formation, especially for sensitive topics, understanding the effects of personalized recommendation beyond accuracy has become essential in today's digital society. In this work, we present \emph{\nemig}, a bilingual news collection on the topic of migration, and corresponding rich user data. 
  In comparison to existing news recommendation datasets, which comprise a large variety of monolingual news, \nemig{} covers articles on a single controversial topic, published in both Germany and the US. We annotate the sentiment polarization of the articles and the political leanings of the media outlets, in addition to extracting subtopics and named entities disambiguated through Wikidata. These features can be used to analyze the effects of algorithmic news curation beyond accuracy-based performance, such as recommender biases and the creation of filter bubbles. We construct domain-specific knowledge graphs from the news text and metadata, thus encoding knowledge-level connections between articles. Importantly, while existing datasets include only click behavior, we collect user socio-demographic and political information in addition to explicit click feedback. We demonstrate the utility of \nemig{} through experiments on the tasks of news recommenders benchmarking, analysis of biases in recommenders, and news trends analysis. \nemig{} aims to provide a useful resource for the news recommendation community and to foster interdisciplinary research into the multidimensional effects of algorithmic news curation. 
\end{abstract}


\maketitle

\section{Introduction}
\label{sec:intro}

The large volume of news published online each day creates an information overload that exceeds the consumptive capacities of readers. At the same time, the digitalization of news consumption has sparked an unprecedented development of personalized recommendation algorithms, used by online news platforms to process the continuously growing quantities of news and offer readers personalized suggestions. However, models which are optimized to maximize congruity to users' preferences and past behavior, tend to produce recommendations which are highly similar in content to previously read/clicked ones \cite{wu2020sentirec,liu2021interaction}. By filtering out news deemed irrelevant to the user, news recommenders control how and which information is disseminated to the public \cite{pariser2011filter}. Consequently, algorithmic news curation has a large impact on opinion formation and voting behavior \cite{bartels1993messages,dewenter2019can}, as well as the potential to fuel conflicts and polarization \cite{jensen2012political,bakshy2015exposure,boutyline2017social,liu2021interaction,ludwig2022does}, especially when dealing with controversial subjects such as war, climate change, or migration. In addition to the power of recommender systems to shape people's perception of the world, the fundamental role that media plays in today's society as a public forum \cite{balkin2017free,helberger2019democratic}, have deemed analyzing and understanding the implications of news curation algorithms necessary for modern democracies. 

While plethora of recommendation algorithms have been proposed in recent years, resources available to analyze their effects beyond accuracy performance and in multilingual scenarios are scarce. The existing body of work exhibits two main shortcomings: 
(1) news datasets for recommendation (i) mostly focus on general or less sensitive topics (e.g., entertainment, fashion, sports) \cite{10.1145/3106426.3109436,WuQCWQLLXGWZ20} and (ii) leave largely unexplored features which are critical for analyzing the algorithmic creation of "filter bubbles" \cite{pariser2011filter} or underlying recommender biases (e.g., sentiment or political orientation); 
(2) news collections for media analysis (e.g., fake news detection \cite{wang2017liar,horne2018sampling}, news narratives analysis \cite{shu2020fakenewsnet,levi2020compres}, news bias detection \cite{farber2020multidimensional,lim2020annotating}) lack user information, and thus, cannot easily be used in a recommendation scenario.

In this paper, we introduce \emph{\nemig{}} a bilingual news collection and knowledge graphs (KGs) about refugees and migration, in German and English, as well as corresponding real-world user data. We collect over 7K, and respectively, 10K articles from German and US news outlets. The articles span a large political spectrum in order to cover various viewpoints on the topic. \textbf{1)} Compared to existing resources, 
\nemig{} (i) targets the same polarizing topic in two languages and (ii) is annotated with the sentiment polarization of articles and political leanings of the media outlets, in addition to subtopics and disambiguated named entities. \textbf{2)} We construct corresponding knowledge graphs (KGs) from the news' text, metadata, and extracted named entities, which we further expand with up to two-hop neighbors from Wikidata of the named entities. We provide the KGs in different variants. \textbf{3)} We collect and publish in an anonymized fashion both explicit feedback and socio-demographic data for 3K users for each of the two datasets. Although our user dataset is small compared to large benchmark news datasets, it contains information about the users' media consumption, political attitudes and interests, personality traits and demographics, which makes the dataset valuable for studying the effects of recommender systems beyond accuracy-based performance, e.g., on political polarization. Furthermore, the user data constitutes a starting point for the generation of synthetic user datasets, which comprise not only click behavior information, but also explicit data about the users' background and preferences.  

We demonstrate the utility of the proposed resource through experiments on different downstream tasks. Firstly, we benchmark several news recommenders on \nemig{}. Secondly, we investigate biases and polarization effects of the benchmarked recommendation models, and evaluate the contribution of various features to the quality of the KG. Lastly, we analyze trends in the news by examining the evolution over time and political orientation of the most frequent entities in our corpora, identifying correlations between the patterns of evolution and worldwide events. 

\nemig{} is available in two versions, for research purposes: i) we release the full news datasets in N-Triples format, under restricted access on Zenodo \cite{iana_andreea_2022_7908392}, given that due to copyright policies, the body of the news can only be provided upon request, ii) \nemig{}, with all features with the exception of the news bodies, is freely available, under a CC BY-NC-SA 4.0 license, in tabular format on GitHub.\footnote{\url{https://github.com/andreeaiana/nemig_nrs/tree/main/data}}
The user data is available in anonymized tabular format on both Zenodo \cite{iana_andreea_2022_7908392} and GitHub.\footnote{\url{https://github.com/andreeaiana/nemig/tree/main/data/user_data}} 

\section{Related Work}
\label{sec:related_work}

\vspace{1.4mm} \noindent \textbf{News Recommendation Benchmarks.}
Various datasets have been constructed for developing and benchmarking news recommender systems (NRS). 
Plista \cite{10.1145/2516641.2516643} comprises a collection of over 70K German articles gathered from 13 news portals, as well as click data for over 14 million users. 
Globo \cite{MoreiraFC18,MoreiraJC19} is a Portuguese dataset for news recommendations collected from {\tt Globo.com}, and contains data for over 300K users and 46K news distributed in 1.2 million sessions.
The Norwegian dataset Adressa \cite{10.1145/3106426.3109436} contains news collected from the Adresseavisen’s news portal and is provided in two variants: a light version with over 11K articles and click data for 561K users, and a larger one with more than 48K articles and clicks for 3 million users. In addition to metadata (e.g., categories, authors), the articles are annotated with named entities. 
MIND \cite{WuQCWQLLXGWZ20}, the most recent large-scale dataset, is constructed from user click logs of Microsoft News. It covers 1 million users and more than 161K news articles in English. Moreover, the dataset is enriched with named entities linked to Wikidata. 
 
\vspace{1.4mm} \noindent \textbf{News Knowledge Graphs.}
Knowledge graphs have been shown to model knowledge-level connections between news that cannot be captured by purely text-based models \cite{wang2018dkn,iana2022survey}. In the context of the Newsreader\footnote{\href{http://www.newsreader-project.eu}{http://www.newsreader-project.eu}} project, focusing on the multilingual processing of news articles, the authors generate an event-centric KG from a multilingual news dataset that describes which events took place, where, when, and who was involved in them \cite{2016kbs,2016jws}. 
News Graph \cite{liu2019news} is a graph constructed specifically for news recommendation. It is generated from a news corpus collected from MSN News containing over 621k articles, 594k news entities, as well as user-item interaction logs. The graph contains news content, user behaviors, and news topic entities enriched with neighboring triples from Microsoft Satori, along with three types of collaborative relations for entities, i.e., co-occurring in the same news, clicked by the same user, and clicked by the same user in the same browsing session. The use of semantic KGs for the production, distribution, and consumption of news is summarized in \cite{10.1145/3543508}.

\vspace{1.4mm} \noindent \textbf{News Datasets for Media Discourse Analysis.}
The advent of news platforms as an ubiquitous means of information for Internet users has established online news as a fundamental source for various media discourse analysis tasks.
For instance, Horne et al. \cite{horne2018sampling} has created a large dataset of political news collected from mainstream and alternative sources, enriched with content-based and social media engagement features, that can be used for news and engagement characterization, news attribution, and content copying, or exploration of news narratives. 
Similarly, Lim et al. \cite{lim2020annotating} and F{\"a}rber et al. \cite{farber2020multidimensional} collected and annotated articles  with different types of bias at the sentence level for analyzing news biases. 
In a related line of work, researchers have designed datasets for monolingual  \cite{wang2017liar,shu2020fakenewsnet} and multilingual cross-domain \cite{shahi2020fakecovid} fake news detection, or for distinguishing between fake and satire news published online \cite{golbeck2018fake,liu2019detection}.

\vspace{1.4mm} \noindent \textbf{Limitations of Current News Resources.} Existing news datasets have several shortcomings. On the one hand, the large monolingual recommendation benchmarks benefit from sizable user interaction data, but fail to explore news (e.g., sentiment orientation, political leaning) or user features (e.g., demographics, political interests) which can be exploited in the analysis of recommender system effects. On the other hand, the news collections for media discourse analysis contain rich annotations (e.g., news biases, social media features, social context, spatio-temporal information), but are ill-equipped to be seamlessly utilized in training recommendation algorithms due to the lack of user feedback data. In contrast to these resources, \nemig{} comprises a (i) bilingual news collection, annotated not only with text and metadata information, but also with subtopic, sentiment and political orientation information, and (ii) implicit user feedback, demographic, and political data. Moreover, we use Wikidata as an open source for entity linking, and not a commercial knowledge base, as done in \cite{liu2019news}. 

\section{The \nemig{} News Corpora}
\label{sec:news_corpora}
 
\subsection{Data Collection}
\begin{figure}[t]
     \centering
     \begin{subfigure}[b]{0.47\textwidth}
         \centering
         \includegraphics[width=\textwidth]{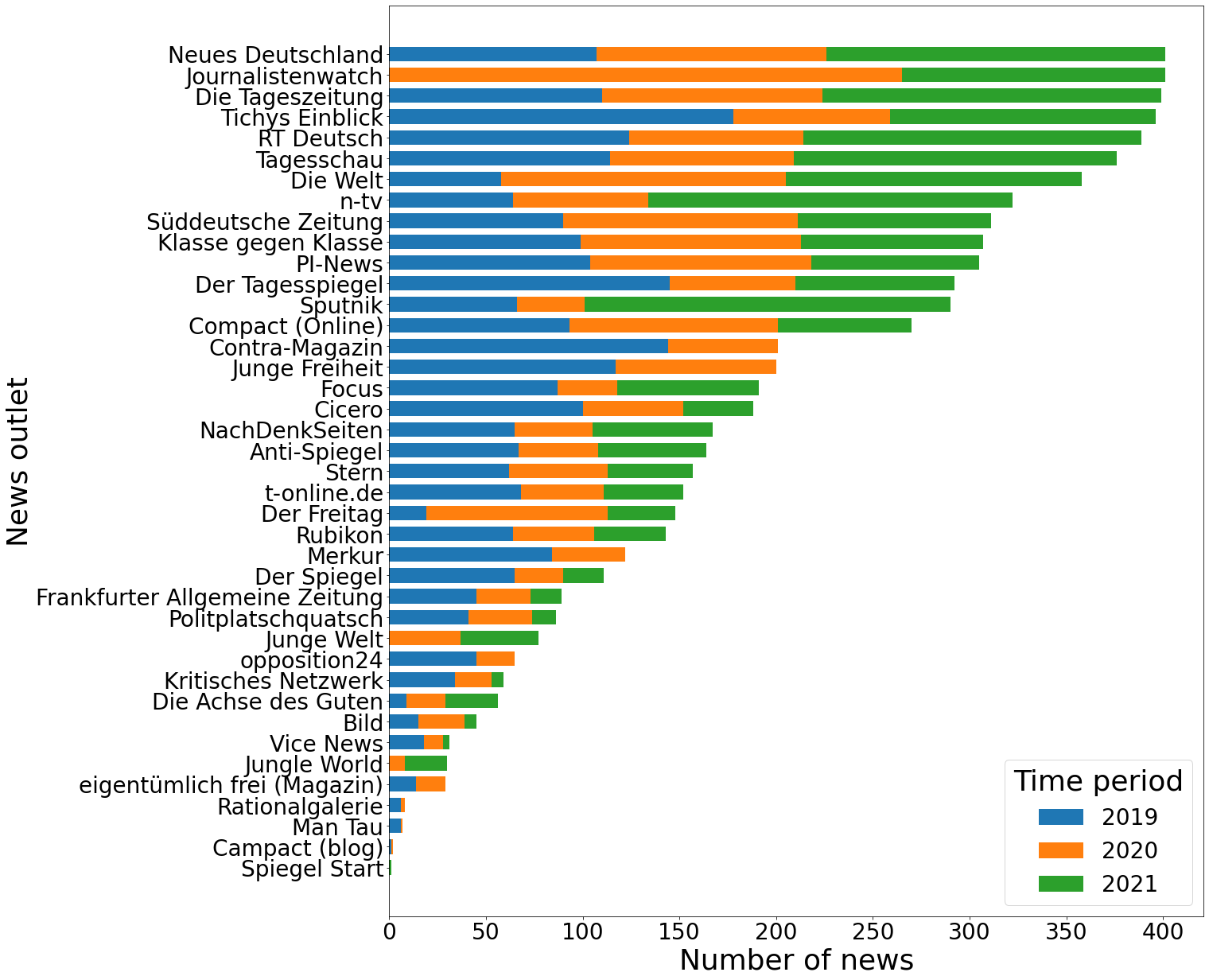}
         \caption{German news corpus.}
         \label{fig:de_corpus_distribution}
     \end{subfigure}
     \hfill
     \begin{subfigure}[b]{0.47\textwidth}
         \centering
         \includegraphics[width=\textwidth]{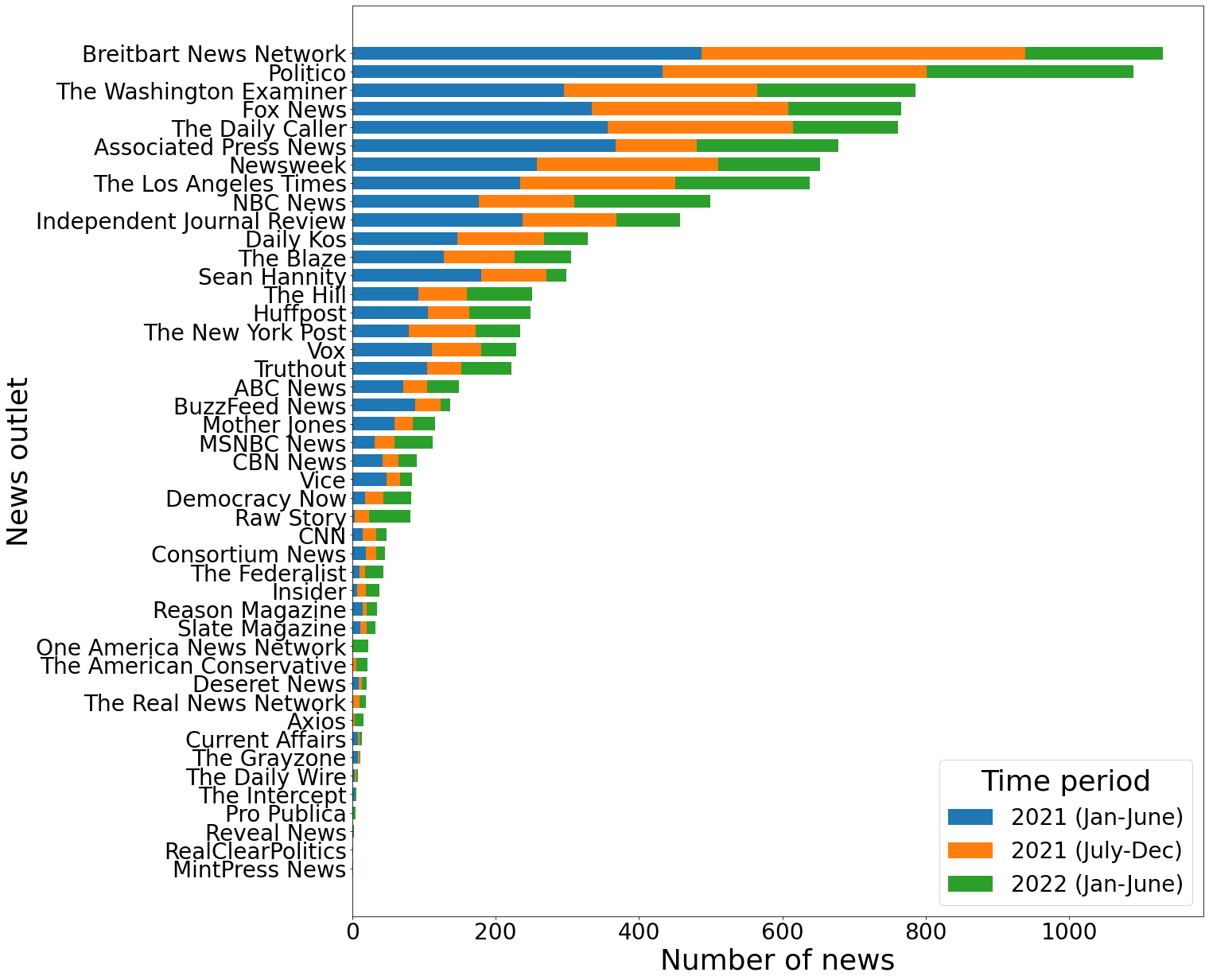}
         \caption{English news corpus.}
         \label{fig:en_corpus_distribution}
     \end{subfigure}
     
    \caption{Distribution of news over time and over outlets.}
    \label{fig:corpus_distribution}
\end{figure}

We construct the news corpora by crawling news articles from 40 German\footnote{The implementation of the crawlers for the German news collection is available at \url{https://github.com/andreeaiana/german-news}}, and respectively 45 US\footnote{The implementation of the crawlers for the English news collection is available at \url{https://github.com/andreeaiana/us-news}}, legacy and alternative media outlets, selected by a team of researchers from the media and communication domain. The news are sampled from the entire political spectrum, such as to cover different viewpoints on the topic. We include German articles published between Jan. 1, 2019 and Dec. 31, 2021\footnote{The data was collected in Dec. 2020 and Jan. 2021, and updated in Jan. 2022.}, and English ones between Jan. 1, 2021 and July 1, 2022\footnote{The data was collected in Aug. 2022.}. For each language, we identify relevant articles using 13 keyword stems representative of the migration topic, such as \textit{flüchtl*}, \textit{asyl*}, or \textit{migration*} for the German outlets, and \textit{refugee*}, \textit{asylum seeker*}, or \textit{migrant*} for the US sources. Moreover, only articles written in either German or English (for the German and English corpora, respectively), with a minimum length of 150 words, and containing at least two keywords stems have been included in the two corpora in order to avoid foreign language articles, disclaimers, advertisements, or reader comments. From the resulting raw datasets, we further excluded duplicates, videos, live tickers (articles continuously updated with news headlines), and outliers\footnote{Outliers were identified by using two standard deviations from the mean.} (articles which are abnormally long or short). The datasets contain 7,346 (German), and respectively 10,814 (English) filtered articles. Figs. \ref{fig:de_corpus_distribution} and \ref{fig:en_corpus_distribution} show the distribution of the articles over time and over media outlets. We observe that the news is unevenly distributed over various publishers, with smaller or niche media outlets being underrepresented. In comparison, with the exception of a few outlets, the articles are relatively evenly distributed over the different time periods.

Each news article contains a title, a body, and if available, an abstract. Additionally, it contains metadata information specifying its provenance (news outlet, URL), publishing and modification dates, authors (persons or organizations), and keywords provided by the media outlet. 
Furthermore, we classify the political leaning of the news outlets. Concretely, we categorize the US sources into three political orientation classes, namely \textit{left}, \textit{center}, and \textit{right}, based on the AllSides Media Bias Chart\footnote{\href{https://www.allsides.com/media-bias/media-bias-chart}{https://www.allsides.com/media-bias/media-bias-chart}}. Since no corresponding classification exists for the German media, the same researcher team classifies the outlets into four political orientation groups: \textit{left}, \textit{center}, \textit{right}, \textit{conspiracy}.\footnote{Disclaimer: We excluded the outlet \textit{Man Tau} from further analysis regarding the political orientation as it cannot be categorized in any of the four groups, but we include the corresponding news in the dataset.} 
We find that the distribution of news outlets over the political orientation classes differs between the German and the English datasets. While the majority of outlets for German are center media, the majority of outlets for English are left. On the article level, however, the distribution for the English dataset is drastically different and similar between the two corpora.

\begin{figure}[t]
    \centering
    \includegraphics[width=\textwidth]{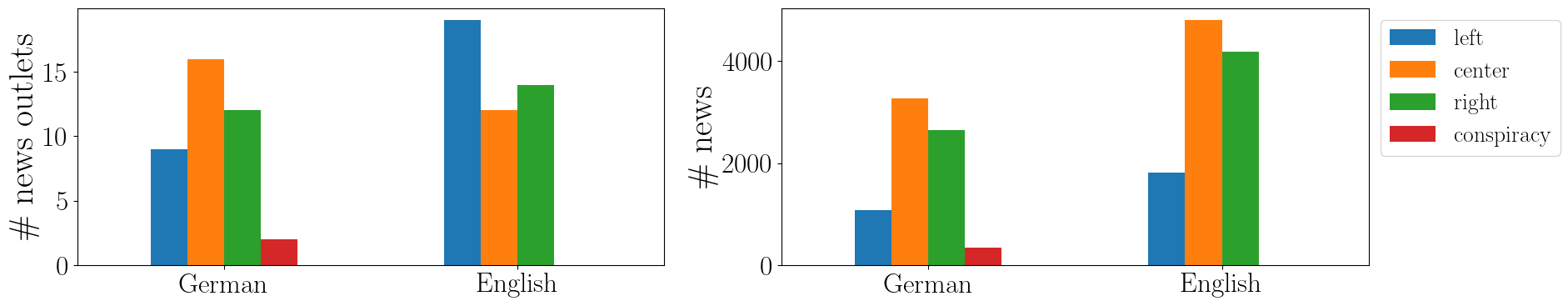}
    \caption{Distribution of outlets and news over political orientation classes.}
    \label{fig:pol_orient}
\end{figure}

\subsection{Data Annotation}
\begin{figure}[t]
    \centering
    \includegraphics[width=0.8\textwidth]{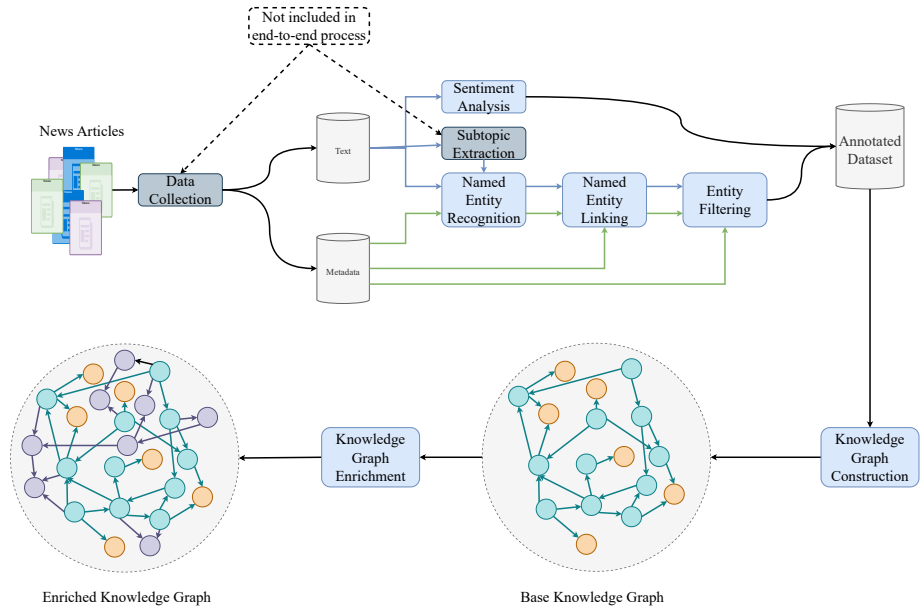}
    \caption{Corpora annotation and knowledge graph construction pipeline. Arrow colors denote the applicability of steps to: all components (black), text features (blue), and metadata (green). Node colors indicate: custom or linked resources (blue), literals (orange), and Wikidata neighboring nodes (purple). Edge colors denote properties in the original KG (blue), or extracted from Wikidata (purple).}
    \label{fig:pipeline}
\end{figure}

Fig. \ref{fig:pipeline} illustrates the corpora annotation and KG construction pipeline.\footnote{The implementation of the annotation and KG construction pipeline, and the intermediary files produced by the annotation process are available at \url{https://github.com/andreeaiana/nemig}}
Given an input dataset, the majority of steps can be run end-to-end.\footnote{The sub-topic extraction step constitutes the only exception, as it requires setting parameters (e.g., threshold for pruning topics) based on intermediary results. We run this process separately and integrate its output within the pipeline.}

\vspace{1.4mm} \noindent \textbf{Sentiment Analysis.}
Sentiment polarization information is leveraged by fairness-aware news recommenders to produce sentiment-diverse or agnostic recommendations \cite{wu2020sentirec,wu2022removing}. Furthermore, such information can be used to investigate whether algorithmic news curation propagates or encourages the sentiment polarization of readers~\cite{mccoy2018polarization,pariser2011filter,cho2020search} or, more generally, the sentiment bias of different news outlets~\cite{gangula2019detecting,lazaridou2016identifying}. We use a multilingual \textit{XLM-R} \cite{conneau2020unsupervised} language model, trained on 198M tweets and fine-tuned for sentiment analysis \cite{barbieri2022xlm}, to classify each article into one of three sentiment classes: positive, neutral, or negative. We perform sentiment classification on the concatenation of the news' title and abstract.\footnote{Note that we use the concatenation of the title and the first sentences of the news body as input when abstracts are not provided.}
Table \ref{tab:corpus_stats} shows the distribution of articles over sentiment classes in the two datasets. For both corpora, only a small number of articles have a positive sentiment. However, while the distribution of articles in the neutral and negative classes is relatively balanced in the German corpus, the English corpus has significantly more negative articles.

\begin{figure}[h]
     \centering
     \begin{subfigure}[b]{\textwidth}
         \centering
         \includegraphics[width=\textwidth]{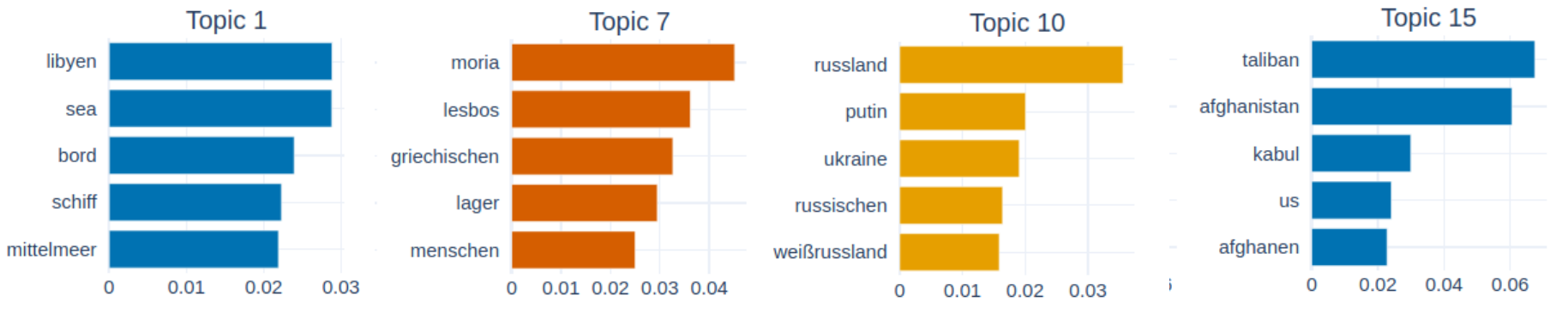}
         \caption{German news corpus.}
         \label{fig:subtopics_de}
     \end{subfigure}
     \vfill
     \begin{subfigure}[b]{\textwidth}
         \centering
         \includegraphics[width=\textwidth]{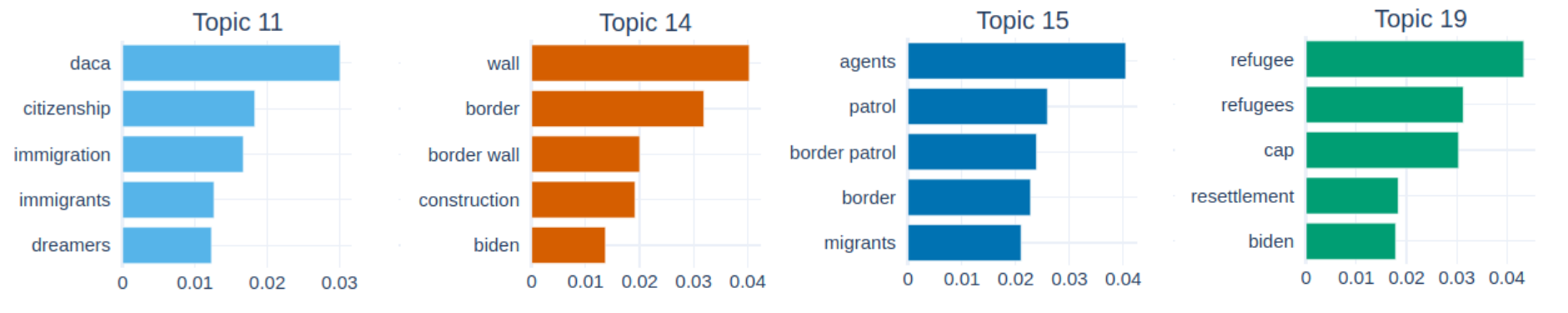}
         \caption{English news corpus.}
         \label{fig:subtopics_en}
     \end{subfigure}
     
    \caption{Examples of words assigned to different sub-topics. The X-axis indicates the c-TF-IDF scores of the words representative of each sampled topic.}
    \label{fig:subtopics}
\end{figure}

\vspace{1.4mm} \noindent \textbf{Sub-topic Modeling.}
\label{para:topic_modelling}
We observe that news cover discourses about the refugee migration from different geographical areas or related to various political events. In order to extract sub-topics from each dataset, we use the neural topic modeling approach proposed in \cite{grootendorst2022bertopic}, with a pre-trained English Sentence Transformer for the English news, and multilingual Sentence Transformer \cite{reimers-2020-multilingual-sentence-bert} for the German ones. We assign labels to each of the resulting topics based on the topmost representative terms of the respective cluster of documents. We group articles that cannot be assigned to any topic appearing in at least 15 news into a separate cluster.
We extract 25 sub-topics from the German corpora, and 40 from the English one, respectively. We find, as shown in Fig. \ref{fig:subtopics}, that the identified sub-topics indicate migration from (e.g., Libya, Ukraine) or to  (e.g., Greece, US) different areas, migration-related political events (e.g., Russian invasion of Ukraine, US army retreating from Afghanistan) or policies (e.g., US immigration and border control laws). 

\vspace{1.4mm} \noindent \textbf{Named Entity Recognition.}
Events are generally described in news by means of named entities (NEs) that indicate what, when, and where something happened, or who was involved in it \cite{li2011personalized}. We extract NEs from both the textual content, and the metadata of articles, and classify them into four classes: persons (PER), organizations (ORG), locations (LOC), and miscellaneous (MISC). We use two \textit{XLM-R} \cite{conneau2020unsupervised} language models, one fine-tuned on the German subset\footnote{\href{https://huggingface.co/xlm-roberta-large-finetuned-conll03-german}{https://huggingface.co/xlm-roberta-large-finetuned-conll03-german}} and the other on the English subset\footnote{\href{https://huggingface.co/xlm-roberta-large-finetuned-conll03-english}{https://huggingface.co/xlm-roberta-large-finetuned-conll03-english}} of the CoNLL03 dataset \cite{sang2003introduction} to perform named entity recognition. 
Concretely, we input the title, abstract, or body of the news after applying sentence segmentation \cite{qi2020stanza} for text components, whereas for metadata, we use author names, lists of keywords from the news outlets, and those used to describe sub-topics. In the former case, we output only the NEs recognized in these original input. In the latter scenario, the output consists not only in NEs extracted by the model but also of the original inputs, where no entity was extracted. In the KG construction step, we include NEs extracted from metadata, as well as other information (e.g., frequent keywords, unrecognized authors) to model connections between articles. 

\begin{table}[t]
    \begin{minipage}{.45\linewidth}
      \caption{News corpora statistics.}
      \centering
      \resizebox{0.8\columnwidth}{!}{
        \begin{tabular}{ll|rr}
            \toprule
            \multicolumn{2}{c}{} 
            & \textbf{German} & \textbf{English} 
            \\ 
            \hline
            
            \multirow{5}{*}{\# News}
            & raw & 8,647 & 12,220 \\
            & filtered & 7,346 & 10,814 \\
            \cdashline{2-4}
            
            & positive & 144 & 590 \\
            & negative & 3,476 & 2,933 \\
            & neutral & 3,726 & 7,291 \\
            \hline

            \multirow{3}{*}{Avg \#words}
            & title & 7.9 & 11.8 \\
            & abstract & 29.4 & 24.9 \\
            & body & 898.4 & 1,013.9 \\
            \bottomrule
            
        \end{tabular}
        \label{tab:corpus_stats}
    }
    \end{minipage}%
    
    \quad
    
    \begin{minipage}{.45\linewidth}
      \centering
        \caption{Statistics of extracted named entities}
        \resizebox{\columnwidth}{!}{
        \begin{tabular}{l|rrr|rrr}
            \toprule
            \multicolumn{1}{c}{} & \multicolumn{3}{c}{\textbf{German}} & \multicolumn{3}{c}{\textbf{English}} \\ 
            \cmidrule(lr){2-4} \cmidrule(lr){5-7}
            
            & Total & Linked & Not linked
            & Total & Linked & Not linked  \\ \hline
            
            Title & 1,075 & 1,075 & 0 & 1,542 & 1,542 & 0 \\
            Abstract & 2,383 & 2,383 & 0 & 2,365 & 2,365 & 0 \\
            Body & 19,683 & 19,683 & 0 & 33,857  & 33,857 & 0 \\ \hdashline
            
            Publishers & 40 & 34 & 6 & 45 & 44 & 1 \\
            Authors & 490 & 193 & 297 & 1,242  & 386 & 856 \\
            Keywords & 4,481 & 2,395 & 2,086 & 5,175 & 2,636 & 2,539 \\
            \bottomrule
        \end{tabular}
        }
        \label{tab:ner_nel_results}
    \end{minipage}  
\end{table}

\vspace{1.4mm} \noindent \textbf{Entity Linking.}
Next, we disambiguate the NEs through named entity linking to Wikidata \cite{vrandevcic2014wikidata} using a multilingual entity linking model \cite{de2022multilingual}, which is based on a sequence-to-sequence architecture \cite{liu2020multilingual} that generates entity names in over 100 languages. 
As input, we use a text sequence containing one named entity annotated with special start and end tokens. The input text sequence is either the title, abstract, or sentence-segmented news body, or for metadata, either the extracted entity from the previous step or the original keyword or author name. For both types of input, the model outputs the linked entity, the corresponding Wikidata QID, the language of the Wikidata entity label, as well as a score. We select Wikidata as the external knowledge base for linkage and graph expansion due to (i)
being open source, (ii) its wide coverage, and (iii) its up-to-dateness \cite{heist2020knowledge}, which is particularly important in the context of news.

\vspace{1.4mm} \noindent \textbf{Entity Filtering.}
The previous step can yield both entities not identified in Wikidata although they exist (e.g., keyword \textit{academia} corresponds to Wikidata QID \textit{Q1211427}), as well as incorrect links. Therefore, we filter incorrectly extracted or linked entities. For NEs originating from textual components, we (1) threshold on the  confidence score of the named entity recognition model; (2) remove entities without a Wikidata page or linked to a Wikimedia disambiguation page; (3) remove entities whose type does not correspond to the entity type in Wikidata based on type-specific properties\footnote{We used the following properties to detect the type of an entity in Wikidata: PER is \textit{instance of} \textit{human}; ORG has \textit{headquarters location}, \textit{inception}, or is \textit{founded by}; LOC has \textit{coordinate location}.}; (4) threshold on the score of the named entity linking model; (5) remove entities whose Wikidata label is in a foreign language and are only observed once in the dataset. For authors and keywords, we use similar filtering pipelines.
Table \ref{tab:ner_nel_results} shows the statistics for the named entity recognition and linking steps.

\begin{figure}[t]
    \centering
    \includegraphics[width=\textwidth]{figures/kg_example.pdf}
    \caption{Example of an article's representation in \nemigkg. The node colors indicate: (i) blank nodes (blue); (ii) disambiguated or custom resources (green); (iii) literals (yellow); (iv) classes (red). The gray boxes denote relation types.}
    \label{fig:kg_example}
\end{figure}

\section{Knowledge Graph Construction}
\label{sec:kg_construction}

Wikidata \cite{vrandevcic2014wikidata} is a large knowledge graph comprising a vast amount of information that is often used to model and integrate common-sense knowledge into recommender systems \cite{guo2020survey,iana2022survey}. Nevertheless, it also contains data which can be irrelevant for domain-specific recommendation tasks. Thus, we use the annotated corpora to build a domain-specific KG, which combines textual content and metadata from news articles with Wikidata triples of the extracted NEs and their \textit{k}-hop neighbors. 

\subsection{Base graph construction}
\nemigkg{} is a KG in which the nodes denote news content and real-world entities (e.g., authors, locations), while the edges indicate different relation types between them. Fig. \ref{fig:kg_example} illustrates an example of an article in \nemigkg.

\vspace{1.4mm} \noindent \textbf{Relations.} 
We model relations using several schemas and vocabularies. Specifically, we represent
edges between articles and their textual content or metadata with \textit{schema.org}\footnote{\url{https://schema.org/}} relationships, and those involving extracted NEs with the \textit{Simple Event Model} \cite{van2011design}. Furthermore, we encode information about miscellaneous NEs with the \textit{schema:mentions} relation, and map the provenance of NEs to the news text with the  \textit{dcterms:isReferencedBy}.\footnote{\url{https://www.dublincore.org/specifications/dublin-core/dcmi-terms/terms/isReferencedBy/}}.
Additionally, we create the relations \texttt{nemig:sentiment} and \texttt{nemig:political\_orientation} to represent edges between news and their sentiment labels, and between media outlets and their political leaning, respectively. 
Lastly, we identify pairs of related news\footnote{Related news are identified using the overlap ratio between pairs of news' bodies.}, i.e., articles with identical titles, overlapping bodies, but different provenance, which we map with the \textit{schema:isBasedOn} relation. We find that such related news are based on one another, with the most recent published one representing an update, extension, or longer discussion of the original article.

\vspace{1.4mm} \noindent \textbf{Nodes.}
\nemigkg{} contains two kinds of nodes: (i) \textit{literals} (e.g., title, dates), and (ii) \textit{resources} which encode NEs extracted from different parts of an article’s content (e.g. persons, locations) or metadata (i.e., publisher, author). We further distinguish the latter into (i) disambiguated entities from Wikidata (denoted as \textit{Wikidata resources}) and (ii) \textit{custom resources} created from entities that could not be identified or linked to Wikidata, but which still encode meaningful information and provide knowledge-level connections between news (e.g., authors not in Wikidata, or frequent keywords). We represent each article in \nemigkg{} with a unique identifier. For each title, abstract, body, topic, sentiment, political leaning, or event mapped in the KG, we create additional nodes, as shown in Fig. \ref{fig:kg_example}. 

\vspace{1.4mm} \noindent \textbf{Enrichment with external information.}
We extend \nemigkg{} with up to \textit{2}-hop neighbors from Wikidata of all \textit{Wikidata resources} already included in our graph to inject additional information from external knowledge bases. Firstly, we extract, for all  \textit{Wikidata resources}, all triples in Wikidata containing another entity as their tail. We ignore literal neighbors. Secondly, we incorporate the matching triples into the graph and retrieve a new neighbors set for the added tail entities. We repeat this process iteratively until we enrich \nemigkg{} with all relevant triples two hops away from the original \textit{Wikidata resources}. We post-process the resulting graph by removing all sink entities, i.e., nodes with a degree smaller than two, which cannot be used to derive further connections between the news as they are leaf nodes or not linked to Wikidata.

\begin{table}[t]
\caption{Statistics of the German and English \nemigkg{}.}
\resizebox{\textwidth}{!}{

   \begin{tabular}{ll|rrrrr|rrrrr}
   \toprule
     \multicolumn{2}{c}{} & \multicolumn{5}{c}{\textbf{German \nemigkg}} & \multicolumn{5}{c}{\textbf{English \nemigkg}}  \\ \cmidrule(lr){3-7} \cmidrule(lr){8-12}
     
     & 
     & Base & Entities & Enriched ($k=1$) & Enriched ($k=2$) & Complete 
     & Base & Entities & Enriched ($k=1$) & Enriched ($k=2$) & Complete\\ 
     \hline     
     
    \multirow{3}{*}{Triples} 
    & Nodes 
    & 89,601 & \cellcolor{Gray} 59,480 & 97,021 & \cellcolor{Gray} 367,514 & 397,635 
    & 134,775 & \cellcolor{Gray} 91,833 & 138,734 & \cellcolor{Gray} 433,965 & 476,907 \\
    
    & Relations 
    & 19 & \cellcolor{Gray} 11 & 768 & \cellcolor{Gray} 1,172 & 1,180 
    & 19 & \cellcolor{Gray} 11 & 828 & \cellcolor{Gray} 1,195 & 1,203 \\
    
    & Triples 
    & 516,948 & \cellcolor{Gray} 458,482 & 821,529 & \cellcolor{Gray} 2,846,849 & 2,905,315 
    & 933,801 & 8\cellcolor{Gray} 47,695 & 1,354,454 & \cellcolor{Gray} 3,599,561 &  3,685,667\\ \hdashline
     
    \multirow{2}{*}{Nodes} 
    & \% resources 
    & 66.38 & \cellcolor{Gray} 100 & 100 & \cellcolor{Gray} 100 & 92.42 
    & 68.14 & \cellcolor{Gray} 100 & 100 & \cellcolor{Gray} 100 & 91.00 \\
    
    & \% literals 
    & 33.62 &\cellcolor{Gray}  0 & 0 & \cellcolor{Gray} 0 & 7.58 
    & 31.86 & \cellcolor{Gray} 0 & 0 & \cellcolor{Gray} 0 & 9.00 \\ \hdashline
     
    \multirow{3}{*}{Resources} 
    & \% blank 
    & 61.28 & \cellcolor{Gray} 61.29 & 37.57 & \cellcolor{Gray} 9.92 & 9.92 
    & 58.48 & \cellcolor{Gray} 58.48 & 38.71 & \cellcolor{Gray} 12.38 & 12.38 \\
    
    & \% custom (not linked) 
    & 4.00 & \cellcolor{Gray} 4.00 & 2.45 &\cellcolor{Gray}  0.65 & 0.65 
    & 3.68 & \cellcolor{Gray} 3.68 & 2.43 & \cellcolor{Gray} 0.78 & 0.78 \\  
    
    & \% Wikidata (linked) 
    & 34.71 & \cellcolor{Gray} 34.71 & 59.97 & \cellcolor{Gray} 89.43 & 89.43 
    & 37.84 & \cellcolor{Gray} 37.84 & 58.86 & \cellcolor{Gray} 86.85 & 86.85\\
    \bottomrule
    \end{tabular}
}
\label{tab:kg_stats}
\end{table}
\subsection{Graph types}

The majority of KG embedding models used in downstream tasks encode only entities and relations, while ignoring literals. Therefore, we create four variants of \nemigkg{} for each of the two datasets, as follows: (1) \textit{base} - contains literals and entities from the corresponding news corpus; (2) \textit{entities} - contains only resource nodes from the \textit{base} graph, and no literals; (3) \textit{enriched entities} - contains only resource nodes and \textit{k}-hop triples from Wikidata; (4) \textit{complete} - represents the \textit{enriched entities} graph with the literals re-added. Table \ref{tab:kg_stats} shows the statistics of the different KGs for both languages.

\section{User Data}
\label{sec:user_data}

We collect user data in terms of (i) explicit click feedback, (ii) demographics, and (iii) political information. For each corpus, we conduct one online user study aimed at measuring the political polarization effects of NRS. The participants are recruited through online-access panels and selected using a quote procedure to create a representative sample of German, and respectively, US Internet users aged 18 to 74. Among the German users, 50.2\% are male and 49.6\% are female, while 44.6\% have an Abitur (passed the secondary school final examinations). 46.2\% of the US users are male and 52.5\% are female, and 33.2\% have at least a high school degree.

In the study, each participant is firstly asked to rate 10 articles as \emph{worthy} or \emph{not to be read further} (i.e., explicit feedback). We use the IDs of the news deemed worthy of reading to build the user \textit{Click History}, which is later used to construct the profile of each user. Afterwards, each participant is shown a list of six news (impressions), for which again a binary rating is recorded. This process was repeated five times. We use the information gathered in the second step to construct the \textit{Impressions} log. Similar to \cite{WuQCWQLLXGWZ20}, we structure the anonymized user behaviors in the format \textit{[impression ID, user ID, Click History, Impression Log]}, where the \textit{Impression Log} contains the IDs of the news shown to the user and the label indicating whether the user clicked on them.  

Before the study's stimulus phase, we collect information regarding the users' media usage (\textit{MED[1-9]}), political attitudes (\textit{POL[1-9]}), and empathy levels (\textit{EMP[1-8]}). After the stimulus, we ask participants questions regarding their socio-demographic status (gender, age, qualification, nationality, born in Germany or the US, parents born in Germany or the US, income), emotions (\textit{EMO[1-14]}), levels of ideological (\textit{IPO[1-12]}), affective (i.e, identification with the main political parties), and perceived (\textit{PPO}[1-6]) polarization, of political participation (\textit{PPA[1-8]}), and of prosocial behavior (\textit{PRO[1-5]}). The resulting user datasets contain information about 3,432 German and 3,000 US users, respectively. The German dataset has a sparsity of only 0.02\%, meaning that out of the news included in the online user study, less than 1\% were never seen by the participants, in either of the two study phases. In contrast, the English dataset is much sparser, with 9.27\% of the news never interacted with by any of the users.
Table \ref{tab:user_study_questions} shows an example of questions used during the user data collection process in the US. Please refer to the project page\footnote{\url{https://github.com/andreeaiana/nemig/tree/main/data/user_data}} or Zenodo \cite{iana_andreea_2022_7908392} for a full set of questions used in both studies, as well as an extensive explanation of each question, and the corresponding answer scale. 

\begin{table}[t]
\caption{A sample of questions used during the online user study conducted in the US for collecting user data.}
\resizebox{\textwidth}{!}{
    \begin{tabular}{ll|lll}
        \toprule
        & \textbf{Variable} 
        & \multicolumn{2}{c}{\textbf{Question}} 
        & \multicolumn{1}{c}{\textbf{Answer Scale}} \\
        \hline

        \multirow{2}{*}{Demographics} 
        & Gender & \multicolumn{2}{l}{Please indicate your gender.} 
        & 0 = male, 1 = female , 2 = other, 3 = no answer 
        \\

         & Nationality 
         & \multicolumn{2}{l}{What is your citizenship?} 
         & \begin{tabular}[c]{@{}l@{}}0 = U.S. citizenship, 1 = U.S. and another non-U.S. citizenship,\\ 2 = Only non-U.S. citizenship, 3 = No Answer\end{tabular} \\
         \hdashline

        \multirow{2}{*}{Empathy} 
        & EMP1 
        & \multirow{2}{*}{\begin{tabular}[c]{@{}l@{}}How strongly do you agree with \\ the following statements?\end{tabular}} 
        
        & When someone else is feeling excited, I tend to get excited too. 
        & \multirow{10}{*}{7-point Likert scale: 1=Strongly disagree to 7=Strongly agree} \\

         & EMP2 
         &  
         & It upsets me to see someone being treated disrespectfully. 
         &  \\
         \cdashline{1-4}

         \multirow{4}{*}{\begin{tabular}[c]{@{}l@{}}Ideological \\ Polarization\end{tabular}} 
         & IPO1 & \multirow{4}{*}{\begin{tabular}[c]{@{}l@{}}In the following, you will see a series of opposing statements. \\ Please indicate how strongly you agree or disagree with the statements. \\ There are no right or  wrong answers.\end{tabular}} 
         & The U.S. should take in more refugees. &  \\
         & IPO2 &  & The U.S. has already taken in too many refugees. &  \\
         & IPO5 &  & Immigrants strive for peaceful cohabitation with U.S.-Americans. &  \\
        & IPO6 &  & Immigrants are hostile toward U.S.-Americans. &  \\
        \cdashline{1-4}

        \multirow{4}{*}{Emotions} & EMO1 & \multirow{4}{*}{\begin{tabular}[c]{@{}l@{}}Which emotions do you feel towards \\ refugees and immigrants in the USA?\end{tabular}} & Anger &  \\
        & EMO2 &  & Fear &  \\
        & EMO9 &  & Compassion &  \\
        & EMO13 &  & Gratitude & \\
        \hdashline

        \multirow{3}{*}{Media Usage} & MED1 & \multirow{3}{*}{\begin{tabular}[c]{@{}l@{}}I receive information about US-American \\ politics via:\end{tabular}} & newspapers and magazines or their websites (e.g. New York Times, The Wallstreet Journal, ...) & \multirow{3}{*}{7-point Likert scale: 1=Never to 7=Very Often} \\
        & MED2 &  & TV networks or their websites (e.g. Fox News, CNN... ) &  \\
        & MED4 &  & Facebook (for political information) &  \\
        \hdashline

        \multirow{2}{*}{\begin{tabular}[c]{@{}l@{}}Perceived \\ Polarization\end{tabular}} & PRO1 & \multirow{2}{*}{\begin{tabular}[c]{@{}l@{}}How strongly do you agree or disagree with \\ the following statements?\end{tabular}} & Democratic and Republican partisans in the U.S. are increasingly hostile to one another. & \multirow{2}{*}{7-point Likert scale:  1=Strongly disagree to 7=Strongly agree} \\
        & PRO3 &  & Democratic and Republican partisans in the U.S. are very polarized. &  \\
        \hdashline

        \multirow{2}{*}{\begin{tabular}[c]{@{}l@{}}Affective \\ Polarization\end{tabular}} & REP & \multirow{2}{*}{\begin{tabular}[c]{@{}l@{}}In the following we would like to know about your party identification. Please mark \\ on the scale how warm or cold you feel towards the respective parties.\end{tabular}} & Republican & \multirow{2}{*}{0 (negative) to 100 (positive)} \\
        & DEM &  & Democrat &  \\
        \hdashline

        Political Scale & POL1 & \multicolumn{2}{l}{Where on the scale would you place your political point of view?} & 11-point Likert Scale: 1 = Left, 6 = Center, 11 = Right \\
        \hdashline

        \multirow{2}{*}{Political Topics} & POL2 & \multirow{2}{*}{\begin{tabular}[c]{@{}l@{}}Please indicate how strongly you agree or disagree \\ with the following statements.\end{tabular}} & I am generally very interested in politics. & \multirow{2}{*}{7-point Likert scale:  1=Strongly disagree to 7=Strongly agree} \\
        & POL3 &  & I regularly inform myself about current political affairs in the U.S.. &  
        \\
        \hdashline

        \multirow{2}{*}{Participation} & PPA1 & \multirow{2}{*}{\begin{tabular}[c]{@{}l@{}}Please indicate how likely it is that you will engage \\ in the following activities in the near future.\end{tabular}} & Participating in an online political discussion on the topic of immigration to the U.S. & \multirow{2}{*}{\begin{tabular}[c]{@{}l@{}}7-point Likert scale:  1=Not likely at all \\ to 7=Very likely\end{tabular}} \\

        & PPA6 &  & Participating in a political demonstration on the topic of immigration to the U.S. &  \\
        \hdashline

        \multirow{2}{*}{\begin{tabular}[c]{@{}l@{}}Prosocial\\ Behavior\end{tabular}} & PRO3 & \multirow{2}{*}{\begin{tabular}[c]{@{}l@{}}Please indicate how strongly you agree or disagree \\ with the following statements.\end{tabular}} & I am willing to make a one-time donation of \_\_\$ for refugees in the U.S. & float number \textgreater{}= 0 \\
        
         & PRO4 &  & How often do you have professional contact (e.g. at work or at school) with immigrants? & 7-point Likert Scale: 1 = Never to 7 = Very often
        \\
         
        \bottomrule
        
        \end{tabular}%
}
\label{tab:user_study_questions}
\end{table}
\section{Experiments}
\label{sec:experiments}

We analyze the performance of different news recommenders on \nemig{} on a range of recommendation tasks. Afterwards,  we perform news trends analysis using the named entities identified in the corpora.  

\subsection{Recommendation Models}

We benchmark various recommendation models, which differ in their news and user encoders.\footnote{The implementation of the news recommenders is available at \url{https://github.com/andreeaiana/nemig_nrs}}
The former converts input features (e.g., title, topics, entities) into a news embedding by contextualizing pretrained word embeddings \cite{pennington2014glove}. The later aggregates the embeddings of the clicked news into a user-level representation. 

\begin{itemize}
    \item \textit{NRMS} \cite{wu2019nrms} embeds news titles with a two-layer encoder consisting of multi-head self-attention \cite{vaswani2017attention}, followed by additive attention \cite{bahdanau2015neural}; a similar architecture is used to learn user representations. 
    
    \item \textit{NAML} \cite{wu2019naml} uses a sequence of convolutional neural network (CNN) \cite{kim-2014-convolutional} and additive attention to learn representations from news titles, and abstracts, and additionally leverages categories through a linear category encoder; its user encoder consists of an additive attention layer.
    
    \item \textit{MINS} \cite{wang2022news} embeds the titles and abstracts of news with the same encoder as NRMS \cite{wu2019nrms}, and categories through a linear embedding layer; it learns user representations with a combination of multi-head self-attention, multi-channel GRU-based recurrent network \cite{cho2014learning}, and additive attention.
    
    \item \textit{CAUM} \cite{qi2022news} combines the text encoder used by NRMS \cite{wu2019nrms} with an entity embedder composed of attention layers in order to learn news representations; it produces candidate-aware user representations with a candidate-aware self-attention network which models long-range dependencies between clicked news, conditioned on the candidate, combined with a candidate-aware CNN that captures short-term user interests from adjacent clicks, also conditioned on the candidate’s content.
    
    \item \textit{DKN} \cite{wang2018dkn} is a knowledge-aware recommender which learns representations from news titles with a knowledge-aware CNN \cite{kim-2014-convolutional} over aligned embeddings of words and entities, and of users with a candidate-aware attention network. 
    
    \item \textit{TANR} \cite{wu2019neural} has the same architecture as NAML \cite{wu2019naml}, but does not use a category embedding layer; additionally, it injects information on topical categories, by jointly optimizing the recommender for content personalization and topic classification. 
    
    \item \textit{SentiDebias} \cite{wu2022removing}, built on the architecture of NRMS \cite{wu2019nrms}, addresses the problem of sentiment debiasing using adversarial learning to reduce the model's sentiment bias (originating from the user data) and generate sentiment-diverse recommendations.

\end{itemize}

We evaluate the aforementioned NRS not only in terms of the standard content personalization performance, but also w.r.t. aspect-based diversity and personalization of results, meaning that we investigate how diverse or faithful the recommendations are to the user's consumed news in terms of aspect $A_p$. Concretely, following \cite{iana2023train}, we define \textit{aspect-based diversification} as the level of uniformity of an aspect's distribution among the recommended news. In contrast, we define \textit{aspect-based personalization} as the level of homogeneity between a user's recommendations and clicked news w.r.t. the distribution of an aspect (e.g., sentiment). We experiment with three aspects: \textit{political leaning}, \textit{topical categories}, and the \textit{sentiment} of news. Note that we use the political leaning of a news outlet as proxy for determining the political leaning of an article originating from that outlet. 

\subsection{Evaluation Metrics}
We use the common metrics AUC, MRR, nDCG@5, and nDCG@10 to report content personalization performance. Following \cite{iana2023train}, we measure aspect-based diversity of recommendations at position $k$ using the normalized entropy of aspects $A_p$'s distribution in the recommendation list:
\begin{equation}
    D_{A_p}@k = - \sum_{j \in A_p} \frac{p(j) \log p(j)}{\log(|A_p|)},
\end{equation}
where $|A_p|$ denotes the number of classes of aspect $A_p$.
We evaluate aspect-based personalization\footnote{Note that perfect aspect-based personalization would imply identical distributions of aspect $A_p$'s in the recommendations list and user history.} with the generalized Jaccard similarity \cite{bonnici2020kullback}:
\begin{equation}
    PS_{A_p}@k = \frac{\sum_{j=1}^{|A_p|} \min(\mathcal{R}_j, \mathcal{H}_j)}{\sum_{j=1}^{|A_p|} \max(\mathcal{R}_j, \mathcal{H}_j)},
\end{equation}
where $R_j$ and $H_j$ represent the probability of a news with class $j$ of aspect $A_p$ to be contained in the recommendations list $\mathcal{R}$, and, respectively, in the user history $\mathcal{H}$. All metrics are bounded to the $[0,1]$ range.

\subsection{Experimental Setup}
We use pre-trained 300-dimensional German\footnote{\href{https://www.deepset.ai/german-word-embeddings}{https://www.deepset.ai/german-word-embeddings}} and English\footnote{\href{https://nlp.stanford.edu/projects/glove/}{https://nlp.stanford.edu/projects/glove/}} GloVe embeddings \cite{pennington2014glove} and 100-dimensional TransD embeddings \cite{ji2015knowledge} pretrained on \nemigkg{} to initialize the word and entity embeddings of the NRS. Note that we use the entity embeddings trained on the variant of \nemigkg{} enriched with 1-hop Wikidata neighbors. Moreover, we use the extracted sub-topics as topical categories in NAML, MINS, and TANR, and embed them with 100-dimensional vectors. Following \cite{ijcai2022infonce}, we sample four negatives per positive example. 
We train all models for 10 epochs, using a batch size of 4.\footnote{We trained each model on a single NVIDIA RTX 2080Ti GPU.}
We optimize with the Adam algorithm \cite{kingma2014adam}, with the learning rate set to 1e-5. We set all other model-specific hyperparameters to optimal values reported in the respective papers. We use 70\% randomly sampled user interactions for training the NRS, 10\% as validation data, and the remaining 20\% as a test set. We repeat each experiment five times with different random seeds and report averages and standard deviations. We normalize the scores to the [0, 100] range.

\subsection{Results and Discussion}

We first discuss the recommendation performance of the aforementioned NRS on \nemig{}. We then analyze whether they are prone to implicit biases in terms of three aspects: \textit{political leaning}, \textit{topical categories}, and \textit{sentiment orientation}. Lastly, we investigate the influence of different input features on the quality of \nemigkg{}.

\vspace{1.4mm} \noindent \textbf{Content Personalization.} 
Table \ref{tab:rec_results} summarizes the results on content personalization for both datasets. We find that, despite the small dataset sizes, the state-of-the-art neural models achieve high predictive performance. We notice a high similarity w.r.t. content personalization performance of NAML and TANR, as well as for NRMS and SentiDebias. Both pairs of models share nearly the same news and user encoder architectures, and differ only in their optimization objectives. These results indicate that secondary optimization goals (i.e., topic classification in the case of TANR, or sentiment debiasing in SentiDebias) have little effect on pure recommendation performance. 

\begin{table}[t]
\caption{Recommendation performance of different NRS. The best results per column are highlighted in bold, the second best are underlined.}
\resizebox{\textwidth}{!}{
    \begin{tabular}{l|rrrr|rrrr}
    \toprule
        \multicolumn{1}{c}{} 
        & \multicolumn{4}{c}{\textbf{German}} 
        & \multicolumn{4}{c}{\textbf{English}} \\ 
        \cmidrule(lr){2-5} \cmidrule(lr){6-9}
        
        Model
        & AUC & MRR & nDCG@3 & nDCG@6 
        & AUC & MRR & nDCG@3 & nDCG@6 
        \\ \hline
        
        NRMS 
        & 57.96$\pm0.66$ 
        & \cellcolor{Gray} 47.83$\pm$0.46  
        & 44.90$\pm$0.65  
        & \cellcolor{Gray} 60.49$\pm$0.35
        
        & 52.40$\pm$0.89 
        & \cellcolor{Gray} 41.46$\pm$0.59 
        & 37.49$\pm$1.07  
        & \cellcolor{Gray} 54.21$\pm$0.53 
        \\
        
        NAML
        & 50.49$\pm$0.13 
        & \cellcolor{Gray} 47.37$\pm$0.84 
        & 44.27$\pm$1.20  
        & \cellcolor{Gray} 60.13$\pm$0.65 
        
        & 50.02$\pm$0.20 
        & \cellcolor{Gray} 41.82$\pm$0.70  
        & 38.14$\pm$0.84  
        & \cellcolor{Gray} 54.49$\pm$0.62 
        \\

        MINS 
        & 57.70$\pm$0.53 
        & \cellcolor{Gray} 47.85$\pm$0.78  
        & 44.96$\pm$0.33  
        & \cellcolor{Gray} 60.50$\pm$0.71 
        
        & 52.91$\pm$0.68 
        & \cellcolor{Gray} \underline{41.96$\pm$1.13}  
        & 38.19$\pm$0.41  
        & \cellcolor{Gray} \underline{54.59$\pm$0.41} 
        \\

        CAUM 
        & \underline{58.18$\pm$0.85} 
        & \cellcolor{Gray} \underline{48.17$\pm$1.08}  
        & \underline{45.27$\pm$0.60}  
        & \cellcolor{Gray} \underline{60.74$\pm$0.55}
        
        & \underline{53.10$\pm$0.78} 
        & \cellcolor{Gray} 41.83$\pm$0.93  
        & 38.16$\pm$1.42  
        & \cellcolor{Gray} 54.50$\pm$0.32
        \\
        
        DKN 
        & \textbf{58.73$\pm$0.39} 
        & \cellcolor{Gray} \textbf{48.20$\pm$0.39}  
        & \textbf{45.48$\pm$0.43}  
        & \cellcolor{Gray} \textbf{60.78$\pm$0.29} 
        
        & \textbf{53.69$\pm$0.66} 
        & \cellcolor{Gray} \textbf{42.14$\pm$0.66}  
        & \textbf{38.57$\pm$0.83}  
        & \cellcolor{Gray} \textbf{54.74$\pm$0.55} 
        \\
        
        TANR 
        & 50.62$\pm$0.14 
        & \cellcolor{Gray} 47.28$\pm$0.44  
        & 44.22$\pm$0.51  
        & \cellcolor{Gray} 60.06$\pm$0.33 
        
        & 50.17$\pm$0.10 
        & \cellcolor{Gray} 41.82$\pm$0.12  
        & \underline{38.26$\pm$0.33}  
        & \cellcolor{Gray} 54.49$\pm$0.17 
        \\

        SentiDebias 
        & 56.80$\pm$0.30 
        & \cellcolor{Gray} 47.27$\pm$0.44  
        & 44.09$\pm$0.18  
        & \cellcolor{Gray} 60.04$\pm$0.44 
        
        & 52.63$\pm$0.90 
        & \cellcolor{Gray} 41.63$\pm$1.34  
        & 37.58$\pm$0.13  
        & \cellcolor{Gray} 54.33$\pm$0.58 
        \\
        
        \bottomrule
        
        \end{tabular}%
}
\label{tab:rec_results}
\end{table}
\begin{table}[t]
\caption{Content personalization and aspect diversity (in terms of topical categories, sentiments, and political leaning) performance of different NRS. The best results per column are highlighted in bold, the second best are underlined.}
\resizebox{\textwidth}{!}{
    \begin{tabular}{l|rrrr|rrrr}
    \toprule
        \multicolumn{1}{c}{} 
        & \multicolumn{4}{c}{\textbf{German}} 
        & \multicolumn{4}{c}{\textbf{English}} \\ 
        \cmidrule(lr){2-5} \cmidrule(lr){6-9}
        
        Model
        & nDCG@3 & D\textsubscript{ctg}@3 & D\textsubscript{snt}@3 & D\textsubscript{pol}@3 
        & nDCG@3 & D\textsubscript{ctg}@3 & D\textsubscript{snt}@3 & D\textsubscript{pol}@3 
        \\ \hline
        
        NRMS 
        & 44.90$\pm$0.65 
        & \cellcolor{Gray} 18.55$\pm$0.25  
        & 33.75$\pm$0.54  
        & \cellcolor{Gray} 30.68$\pm$0.32 
        
        & 37.49$\pm$1.07 
        & \cellcolor{Gray} \underline{22.19$\pm$0.34}  
        & \underline{32.50$\pm$0.27}  
        & \cellcolor{Gray} 26.31$\pm$0.94 
        \\
        
        NAML
        & 44.27$\pm$1.20 
        & \cellcolor{Gray} \textbf{19.31$\pm$0.21}  
        & 33.63$\pm$0.18  
        & \cellcolor{Gray} 30.41$\pm$0.37 
        
        & 38.14$\pm$0.84 
        & \cellcolor{Gray} \textbf{23.89$\pm$0.65}  
        & 32.08$\pm$0.64  
        & \cellcolor{Gray} 25.66$\pm$1.24 
        \\

        MINS 
        & 44.96$\pm$0.33 
        & \cellcolor{Gray} 18.18$\pm$0.44  
        & 33.79$\pm$0.34  
        & \cellcolor{Gray} 30.10$\pm$0.44 
        
        & 38.19$\pm$0.41 
        & \cellcolor{Gray} 21.94$\pm$1.11  
        & 32.34$\pm$0.16  
        & \cellcolor{Gray} 25.22$\pm$1.11 
        \\
        
        CAUM 
        & \underline{45.27$\pm$0.60}
        & \cellcolor{Gray} 18.09$\pm$0.33  
        & 33.13$\pm$0.43  
        & \cellcolor{Gray} 30.42$\pm$0.33 
        
        & 38.16$\pm$1.42 
        & \cellcolor{Gray} 21.90$\pm$0.63  
        & 32.09$\pm$0.26  
        & \cellcolor{Gray} 25.90$\pm$0.63 
        \\
        
        DKN 
        & \textbf{45.48$\pm$0.43} 
        & \cellcolor{Gray} 18.46$\pm$0.16  
        & \underline{33.79$\pm$0.44}  
        & \cellcolor{Gray} \underline{30.92$\pm$0.59} 
        
        & \textbf{38.57$\pm$0.83} 
        & \cellcolor{Gray} 22.04$\pm$0.40  
        & 32.46$\pm$0.66  
        & \cellcolor{Gray} \textbf{26.58$\pm$0.90} 
        \\

        TANR 
        & 44.22$\pm$0.51 
        & \cellcolor{Gray} 18.52$\pm$0.18  
        & 33.70$\pm$0.36  
        & \cellcolor{Gray} 30.39$\pm$0.38 
        
        & \underline{38.26$\pm$0.33} 
        & \cellcolor{Gray} 21.59$\pm$0.19  
        & 32.41$\pm$0.68  
        & \cellcolor{Gray} \underline{26.47$\pm$0.82} 
        \\

        SentiDebias 
        & 44.09$\pm$0.18 
        & \cellcolor{Gray} \underline{18.66$\pm$0.30}  
        & \textbf{33.84$\pm$0.34}  
        & \cellcolor{Gray} \textbf{30.99$\pm$0.30} 
        
        & 37.58$\pm$0.13 
        & \cellcolor{Gray} 22.10$\pm$1.11  
        & \textbf{33.26$\pm$0.34}  
        & \cellcolor{Gray} 25.81$\pm$1.11 
        \\
        
        \bottomrule
        
        \end{tabular}%
}
\label{tab:div_results}
\end{table}

Generally, we find that the two knowledge-aware models, DKN and CAUM, outperform all the other recommenders. This suggests that injecting additional external knowledge into the model can help detect additional, knowledge-level connections between news, thus resulting in better news representations. The results are partly at odds with the findings of \cite{iana2023simplifying}, which observed that while CAUM indeed outperforms other models with candidate-agnostic user encoders, DKN underperforms. However, \cite{iana2023simplifying} conducted experiments on the MIND dataset \cite{WuQCWQLLXGWZ20}, which is magnitudes larger than \nemig{}. This points to the fact that both knowledge-awareness and candidate-awareness user modeling might be particularly beneficial in scenarios where little training data is available. Lastly, we observe that the quality of content personalization is higher for the German than for the English dataset. We hypothesize that this is due to the higher sparsity of the English dataset, in which significantly more news have never been seen by the users.

\vspace{1.4mm} \noindent \textbf{Aspect-based Diversity and Personalization.}
Next, we analyze the diversity and personalization of results, using topical categories (ctg), sentiment (snt), and political leaning (pol) as aspects. Concretely, we report the performance for content personalization(nDCG@$k$), aspect-based diversity (D\textsubscript{ctg} for topical categories, D\textsubscript{snt} for sentiment polarization, and D\textsubscript{pol} for political leaning) and aspect-based personalization (PS\textsubscript{ctg} for topical categories, PS\textsubscript{snt} for sentiment polarization, and PS\textsubscript{pol} for political leaning). We report results for $k=3$. 

Table \ref{tab:div_results} summarizes the results on content personalization and aspect diversity. As expected, SentiDebias \cite{wu2022removing} achieves the highest sentiment diversity on both datasets. NAML \cite{wu2019naml} outperforms, by a relatively large margin, the other models in terms of category diversity of results. These results are somewhat surprising and at odds with those obtained on other datasets \cite{iana2023train}, as NAML leverages information on topical categories to customize results to the users' preferences, and not to address diversity. Furthermore, compared to the first two aspects, we find that all recommenders perform nearly identical w.r.t. degree of political diversification of recommendations. This could be due to the fact that none of the models explicitly target political diversification.

Aspect-based diversity is tightly correlated with aspect-based personalization \cite{iana2023train}. More specifically, higher levels of diversity come at the cost of personalization, as can be seen in Table \ref{tab:pers_results}, which illustrates the results on content and aspect-based personalization. We observe that categorical and political personalization are much more aligned with content personalization performance, then with the sentiment of news, as discussed also in \cite{iana2023train}. We find this to be intuitive, as users tend to choose which news to read based on their topics/categories of interest, as well as political preferences, and not on their sentiment polarization.

An earlier version of the German dataset~\cite{andreea_iana_2022_5913171} has been used in~\cite{alam2022towards} and~\cite{ludwig2023divided} to examine sentiment and stance recommender bias, and to investigate polarization and filter bubble creation through online studies, respectively. While in this study we have analyzed political biases of recommenders only from the perspective of users' click histories, in the future their political information can be further explored in conjunction with the models' predictions to understand whether any existing biases are reinforced or amplified by the algorithms. Similarly, \nemig{} can be used in future works to analyze these trends also in multilingual or cross-lingual recommendation scenarios, as it comprises information on the same topic in both German and English.

\begin{table}[t]
\caption{Content and aspect personalization (in terms of topical categories, sentiments, and political leaning) performance of different NRS. The best results per column are highlighted in bold, the second best are underlined.}
\resizebox{\textwidth}{!}{
    \begin{tabular}{l|rrrr|rrrr}
    \toprule
        \multicolumn{1}{c}{} 
        & \multicolumn{4}{c}{\textbf{German}} 
        & \multicolumn{4}{c}{\textbf{English}} \\ 
        \cmidrule(lr){2-5} \cmidrule(lr){6-9}
        
        Model
        & nDCG@3 & PS\textsubscript{ctg}@3 & PS\textsubscript{snt}@3 & PS\textsubscript{pol}@3 
        & nDCG@3 & PS\textsubscript{ctg}@3 & PS\textsubscript{snt}@3 & PS\textsubscript{pol}@3 
        \\ \hline
        
        NRMS 
        & 44.90$\pm$0.65 
        & \cellcolor{Gray} 20.72$\pm$0.45
        & \textbf{42.66$\pm$0.30}  
        & \cellcolor{Gray} 36.55$\pm$0.09
        
        & 37.49$\pm$1.07 
        & \cellcolor{Gray} 18.48$\pm$0.18  
        & 41.38$\pm$0.40  
        & \cellcolor{Gray} 40.66$\pm$0.22 
        \\
        
        NAML
        & 44.27$\pm$1.20 
        & \cellcolor{Gray} 20.42$\pm$0.31  
        & 42.59$\pm$0.25  
        & \cellcolor{Gray} 36.23$\pm$0.32 
        
        & 38.14$\pm$0.84 
        & \cellcolor{Gray} 18.24$\pm$0.81  
        & \textbf{41.43$\pm$0.19}  
        & \cellcolor{Gray} \textbf{41.17$\pm$0.30} 
        \\

        MINS 
        & 44.96$\pm$0.33 
        & \cellcolor{Gray} \textbf{21.10$\pm$0.60}  
        & 42.56$\pm$0.34  
        & \cellcolor{Gray} \textbf{36.74$\pm$0.27} 
        
        & 38.19$\pm$0.41 
        & \cellcolor{Gray} \underline{18.83$\pm$0.41}  
        & 41.20$\pm$0.16  
        & \cellcolor{Gray} \underline{41.15$\pm$0.21} 
        \\

        CAUM 
        & \underline{45.27$\pm$0.60}
        & \cellcolor{Gray} 20.85$\pm$0.79  
        & 42.48$\pm$0.43  
        & \cellcolor{Gray} \underline{36.59$\pm$0.36} 
        
        & 38.16$\pm$1.42 
        & \cellcolor{Gray} 18.67$\pm$1.45  
        & \underline{41.41$\pm$0.26}  
        & \cellcolor{Gray} 40.95$\pm$0.41 
        \\
        
        DKN 
        & \textbf{45.48$\pm$0.43} 
        & \cellcolor{Gray} \underline{20.95$\pm$0.35}  
        & 42.51$\pm$0.36  
        & \cellcolor{Gray} 36.08$\pm$0.24 
        
        & \textbf{38.57$\pm$0.83} 
        & \cellcolor{Gray} 18.56$\pm$0.35  
        & 41.25$\pm$0.14  
        & \cellcolor{Gray} 40.70$\pm$0.15 
        \\
        
        TANR 
        & 44.22$\pm$0.51 
        & \cellcolor{Gray} 20.88$\pm$0.24  
        & \underline{42.64$\pm$0.19}  
        & \cellcolor{Gray} 36.42$\pm$0.34 
        
        & \underline{38.26$\pm$0.33} 
        & \cellcolor{Gray} \textbf{19.10$\pm$0.29}  
        & 41.27$\pm$0.24  
        & \cellcolor{Gray} 40.98$\pm$0.20 
        \\

        SentiDebias 
        & 44.09$\pm$0.18 
        & \cellcolor{Gray} 20.50$\pm$0.39  
        & 42.64$\pm$0.34  
        & \cellcolor{Gray} 36.40$\pm$0.24 
        
        & 37.58$\pm$0.13 
        & \cellcolor{Gray} 18.57$\pm$0.33  
        & 41.29$\pm$0.34  
        & \cellcolor{Gray} 40.84$\pm$0.20 
        \\

        \bottomrule
        
        \end{tabular}%
}
\label{tab:pers_results}
\end{table}
\begin{table}[t]
\caption{Content personalization and aspect-based diversity (in terms of topical categories, sentiment, and political leaning) performance of DKN with different variants of \nemigkg{}.}
\resizebox{\textwidth}{!}{
    \begin{tabular}{l|rrrrr|rrrrr}
    \toprule
        \multicolumn{1}{c}{} & \multicolumn{5}{c}{\textbf{German}} & \multicolumn{5}{c}{\textbf{English}} \\ \cmidrule(lr){2-6} \cmidrule(lr){7-11}
        Model 
        & AUC & nDCG@3 
        & D\textsubscript{ctg}@3 & D\textsubscript{snt}@3 & D\textsubscript{pol}@3 
        
        & AUC & nDCG@3 
        & D\textsubscript{ctg}@3 & D\textsubscript{snt}@3 & D\textsubscript{pol}@3 
        \\ \hline
        
        content entities 
        & 58.72$\pm$0.45 
        & \cellcolor{Gray} 45.57$\pm$0.44  
        & 18.44$\pm$0.16  
        &  \cellcolor{Gray} 33.67$\pm$0.38 
        & 31.17$\pm$0.56 

        & 53.67$\pm$0.63 
        & \cellcolor{Gray} 38.60$\pm$1.01  
        & 21.99$\pm$0.45  
        &  \cellcolor{Gray} 32.33$\pm$0.34  
        & 26.71$\pm$0.75 
        \\

        + topical categories
        & 58.82$\pm$0.23 
        & \cellcolor{Gray} 45.74$\pm$0.31  
        & 18.43$\pm$0.09  
        &  \cellcolor{Gray} 33.83$\pm$0.39  
        & 30.93$\pm$0.64 

        & 53.67$\pm$0.50 
        & \cellcolor{Gray} 38.68$\pm$0.72  
        & 22.02$\pm$0.38  
        &  \cellcolor{Gray} 32.20$\pm$0.45  
        & 26.47$\pm$1.12 
        \\
        
        + sentiment
        & 58.81$\pm$0.26 
        & \cellcolor{Gray} 45.69$\pm$0.38  
        & 18.48$\pm$0.08  
        & \cellcolor{Gray} 33.77$\pm$0.44  
        & 30.98$\pm$0.65 

        & 53.70$\pm$0.53 
        & \cellcolor{Gray} 38.53$\pm$0.73  
        & 21.95$\pm$0.37  
        &  \cellcolor{Gray} 32.17$\pm$0.18  
        & 26.89$\pm$0.72 
        \\
        
        + political leaning 
        & 58.73$\pm$0.42 
        & \cellcolor{Gray} 45.55$\pm$0.43  
        & 18.45$\pm$0.16  
        &  \cellcolor{Gray} 33.68$\pm$0.47  
        & 31.00$\pm$0.45 

        & 53.73$\pm$0.72 
        & \cellcolor{Gray} 38.66$\pm$0.79  
        & 22.00$\pm$0.46  
        &  \cellcolor{Gray} 32.28$\pm$0.36  
        & 26.74$\pm$0.52 
        \\ \hdashline
        
        + 1-hop neighbors 
        & 58.82$\pm$0.33 
        & \cellcolor{Gray} 45.74$\pm$0.46  
        & 18.50$\pm$0.09  
        & \cellcolor{Gray} 33.77$\pm$0.40  
        & 31.20$\pm$0.50

        & 53.78$\pm$0.44 
        & \cellcolor{Gray} 38.76$\pm$0.65  
        & 21.95$\pm$0.37  
        & \cellcolor{Gray} 32.16$\pm$0.25  
        & 26.60$\pm$0.59 
        \\

        + 2-hop neighbors 
        & 58.73$\pm$0.39 
        & \cellcolor{Gray} 45.48$\pm$0.43  
        & 18.46$\pm$0.16  
        & \cellcolor{Gray} 33.79$\pm$0.44  
        & 30.92$\pm$0.59 

        & 53.69$\pm$0.66
        & \cellcolor{Gray} 38.57$\pm$0.83  
        & 22.04$\pm$0.40  
        & \cellcolor{Gray} 32.46$\pm$0.66  
        & 26.58$\pm$0.90 
        \\
        \bottomrule
        
        \end{tabular}%
}
\label{tab:ablation_results}
\end{table}

\vspace{1.4mm} \noindent \textbf{Knowledge Graph Ablations.}
The type of data contained in the knowledge graph is paramount for training high quality knowledge graph embeddings, which are further exploited by knowledge-aware recommenders to improve the accuracy of predictions. Thus, we analyze the impact of input features on \nemigkg{}. Specifically, we pre-train entity embeddings on different versions of \nemigkg{} using TransD \cite{ji2015knowledge}, and study the recommendation and aspect-based diversity performance of DKN using these embeddings. We summarize the results in Table \ref{tab:ablation_results} in terms of content personalization (nDCG@$k$), and aspect-based diversity for topical categories (D\textsubscript{ctg}), sentiment (D\textsubscript{snt}), and political leaning (D\textsubscript{pol}).
Adding topical categories, sentiment and political information in \nemigkg{} increases the diversity of recommendations w.r.t. these aspects, while having minor effects on accuracy-based performance. We find that extending \nemigkg{} with $k$-hop neighbors from Wikidata is beneficial only up to $k=1$ hops. This indicates that a certain degree of contextualization of named entities with general knowledge is helpful when the data size is small. However, injecting too much external knowledge  (i.e., $k=2$) can dilute the original information and have detrimental effects on the recommender's downstream performance.

\begin{figure}[t]
    \centering
    \includegraphics[width=\textwidth]{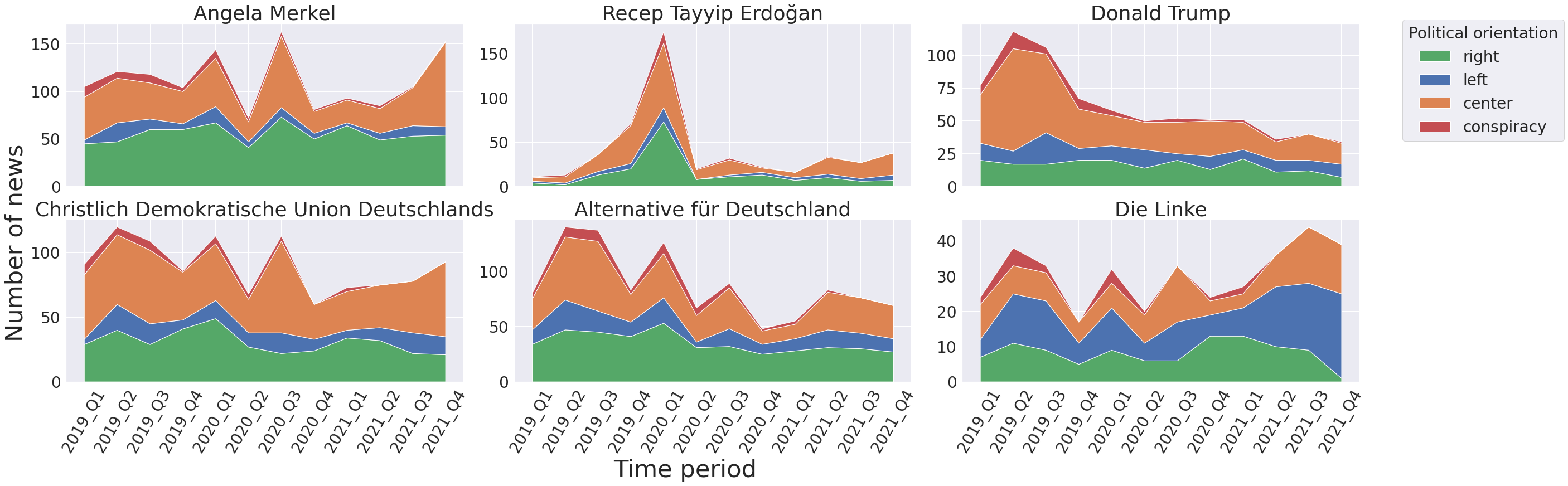}
    \caption{Evolution over time and political orientation of a sample of the top-20 most popular entities from the German dataset. The first row contains entities of type PER, and the second of type ORG.
    }
    \label{fig:news_trends_de}
\end{figure}

\begin{figure}[t]
    \centering
    \includegraphics[width=\textwidth]{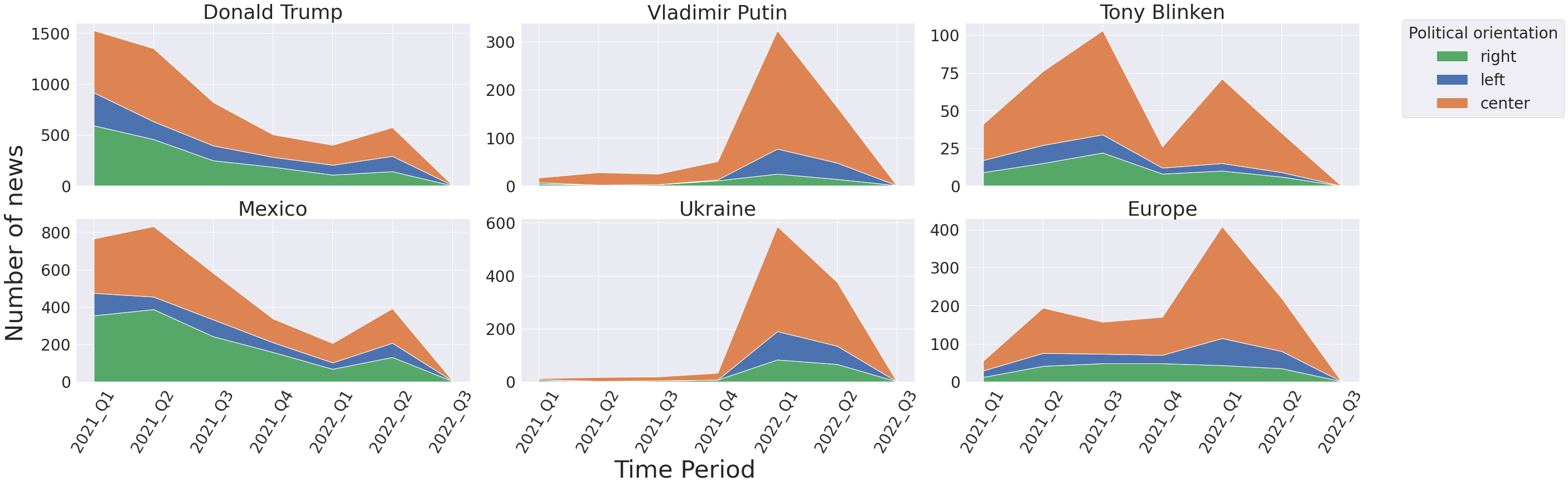}
    \caption{Evolution over time and political orientation of a sample of the top-20 most popular entities from the English dataset. The first row contains entities of type PER, and the second of type LOC.
    }
    \label{fig:news_trends_en}
\end{figure}

\subsection{News Trends Analysis}
\label{subsec:news_trends}

Lastly, we demonstrate how \nemig{} can be used to analyze trends and correlations between entities used in the news discourse and events. Concretely, we analyze the evolution of named entities overtime and political groups. Figs. \ref{fig:news_trends_de}-\ref{fig:news_trends_en} show the evolution over time and political orientation of the source outlets of some of the top-20 most frequent entities in our corpora. We find that generally the most frequently sampled entities are covered predominantly in center media, which is also the source of the majority of news in both corpora. Notable exceptions in the German corpus (Fig. \ref{fig:news_trends_de}) are \textit{Angela Merkel}, appearing equally or more often in right-wing media, and \textit{Die Linke}, with a more equal distribution among all political classes, although left-wing coverage dominates the discourse. The evolution of entities over time reveals correlations with major events. In terms of organizations, German news report most frequently about the main political parties, whose evolution coincides with elections on the federal and European level, or about central political figures in international politics (e.g., the coverage of \textit{Recep Tayyip Erdoğan} appears to be potentially correlated with the Turkish offensive in north-eastern Syria). In contrast to German media, the representation of the top entities from the English corpus in the left US media is generally lower.

\section{Conclusion}
\label{sec:conclusion}
We introduced \nemig{}, a bilingual news collection and knowledge graphs on the topic of migration, along with user data, including demographic and political information. The news is collected from mainstream and alternative German and US media outlets spanning a wide political scale. We extract sub-topics and named entities linked to Wikidata from the news text and metadata. In contrast to existing datasets, we provide sentiment and political annotations, and use Wikidata for incorporating background knowledge. We hope that this resource will inspire future research on analyzing the multidimensional implications of news curation algorithms, in both monolingual and cross-lingual settings.

\begin{acks}
    This work has been conducted in the ReNewRS project, which is  funded by the Baden-Württemberg Stiftung in the Responsible Artificial Intelligence program. The authors would also like to thank Julia Wildgans at the University of Mannheim for her legal advice on data handling.
\end{acks}

\bibliographystyle{ACM-Reference-Format}
\bibliography{references}

\end{document}